\documentclass[aps,prd,preprint,showkeys,showpacs,unsortedaddress,superscriptaddress]{revtex4-1}
\usepackage{amsmath, amsfonts, amssymb, mathrsfs,float}
\usepackage{graphicx}
\usepackage{bm,color}
\newcommand{\nn}{\nonumber}

\newcommand{\sss}{\scriptscriptstyle}
\allowdisplaybreaks

\begin{document}

\title{Post-Newtonian light propagation in Kerr-Newman spacetime}
\author{Chunhua Jiang}%
\affiliation{School of Mathematics and Physics, University of South China, Hengyang, 421001, China}
\author{Wenbin Lin}
\email{To whom all correspondence should be addressed (lwb@usc.edu.cn).}
\affiliation{School of Mathematics and Physics, University of South China, Hengyang, 421001, China}
\affiliation{School of Physical Science and Technology, Southwest Jiaotong University, Chengdu, 610031, China}

\begin{abstract}
The second-order post-Newtonian solution for the light propagation in the field of Kerr-Newman black hole is achieved via an iterative method. Based on this result, we further obtain the second-order post-Newtonian light deflection in Kerr-Newman spacetime, which is formulated in an united form for any arbitrarily incident directions. All results are exhibited in the coordinate system constituted by the initial light-direction vector, the impact vector, and their cross-product.
\end{abstract}

\maketitle

\section{introduction}

Einstein's theory of general relativity is now widely accepted as the most satisfactory theory of gravitation. The motion of test particles including photon is a core problem for general relativity.
In the past decades hampered by the limitations of instruments, the measurements for testing relativistic gravity were up to only the first-order post-Newtonian (1PN) precision.
However, todays astrometric missions, e.g., GAIA~\cite{Turyshev2009}, have reached a level of a few micro-arcseconds ($\mu$ arcsec) in angular observations of light. These projects may measure the effects of relativistic gravity on the light propagation at the second-order post-Newtonian (2PN) level. Therefore, it is important to obtain the analytical solution for the light propagation to this accuracy.

{\color{black}The effects on the light propagation have been explored within several investigations, e.g., the 1PN and 1.5PN effects of the mass-monopole, spin-dipole and mass-quadrupole of gravitational bodies in motion~\cite{Klioner1991,KlionerKopeikin1992}, the post-Minkowskian (1PM) effects of mass-multipoles and spin-multipoles of gravitational sources at rest~\cite{Kopeikin1997,KopeikinKorobkovPolnarev2006,Poncin-LafitteTeyssandier2008}, the 1PM effects of pointlike bodies in motion~\cite{KopeikinSchafer1999,KopeikinMashhoon2002,Klioner2003,KlionerPeip2003}, and the 1PN and 1.5PN effects of both motion and mass-multipoles and spin-multipoles of sources~\cite{Zschocke2015,SoffelHan2015,Zschocke2016}. Furthermore, the canonical studies on the 2PN contributions to the light deflection and time delay were performed in the early of 1980s or even earlier~\cite{EpsteinShapiro1980,FischbachFreeman1980,RichterMatzner1982,RichterMatzner1983,Will2014}. Recently, based on the method of time transfer functions~\cite{LinetTeyssandier2002,TeyssandierPoncin-Lafitte2008},
the formula for the light travelling time in the field of a spherically symmetric and static spacetime has been achieved up to the fifth order of gravitational constant~\cite{LinetTeyssandier2016}.}

{\color{black}Due to their strong gravitational field, the higher-order PN effects caused by the black holes on the passing-by light  might be detected in future.} The light propagation in the field of Schwarzschild black hole has been studied extensively~\cite{Cadez2005,Klioner2010,Bozza2010,Kostic2012,Gibbons2012,Munoz2014}. For the case of Kerr black hole, the trajectory of the light is not confined to a plane in general, except for the light being in the equatorial plane which is the perpendicular to the gravitational source's rotational axis. For simplicity, many analytical works focus on the propagation of light in or close to the equatorial plane~\cite{Bozza2003,Edery2006,IyerHansen2009,Barlow2017}.

Kerr-Newman black hole is the most general classic black hole. The expression for the equatorial deflection of light up to the fourth-order level has been obtained~\cite{ChakrabortySen2015}. {\color{black}In general, similar to the case of Kerr black hole, the light propagation in the field of Kerr-Newman black hole does not proceed in a plane either}. Although the conditions for the orbital and vortical motions of photon outside the equatorial plane have been studied~\cite{CalvaniTurolla1981}, so far for the general situations, the analytical solutions to the trajectory and velocity of light in Kerr-Newman spacetime have not been reported yet.

In this work, we consider the problem of the light with an arbitrary direction travelling in the gravitational field of Kerr-Newman black hole. We follow Will's method~\cite{Will1981} to evaluate differential equations for the null geodesics and take a step forward up to the next higher order, and in the process develop a simple iterative method that allow the trajectory and velocity of light to be calculated to arbitrary PN orders. On the other hand, for any non-spherically-symmetry gravitational system, the light propagation is not confined in a plane, and its deflection can not be characterized by a single deflection angle completely. For this purpose, we also introduce a coordinate system with which the light trajectory, velocity and deflection can be described conveniently.

The present paper is organized as follows. Section \ref{sec:2nd} gives the 2PN dynamics equations for the photon in Kerr-Newman spacetime. In Section \ref{sec:3rd} we present the 2PN solutions to the trajectory and velocity of light travelling in the field of Kerr-Newman black hole. Section \ref{sec:4th} gives the 2PN light deflection in Kerr-Newman spacetime. The summary is given in Section \ref{sec:summary}.

\section{The 2PN dynamics equations of photon}\label{sec:2nd}

The spacetime for a constantly rotating charged black hole is referred to as Kerr-Newman spacetime. {\color{black}In harmonic coordinates}, the metric of Kerr-Newman black hole in the 2PN approximation can be written as \cite{LinJiang2014}
\begin{eqnarray}
g_{00}&=& -1 +\frac{2m}{r} -\frac{2m^2+q^2}{r^2}~,\label{eq:metric-1nd}\\
g_{0i}&=& \zeta^i~,\label{eq:metric-2nd}\\
g_{ij}&=&\Big(1 +\frac{2m}{r} + \frac{m^2}{r^2}\Big)\delta_{ij} + \frac{(m^2-q^2)x^i x^j}{r^4}~,\label{eq:metric-3nd}
\end{eqnarray}
where $m,~q,~\bm{J}$ are Kerr-Newman black hole's mass and electric charge respectively, $r\equiv |\bm{x}|$ with $\bm{x}\equiv(x^1, x^2, x^3)$ {\color{black}denoting the position vector of the field point}, $\zeta^i$ is the $i$-th component of the vector potential $\bm{\zeta} \equiv 2 (\bm{x}\times\bm{J})/r^3$ due to the gravitational source's rotation, with $\bm{J}$ being the angular momentum of Kerr-Newman black hole. We use the geometrized units (the gravitational constant $G$ and the light speed $c$ in vacuum are set as $1$) and the metric signature ($-+++$) with Greek indices running from 0--3 and Latin indices from 1--3.

The motion of photon in the gravitational field can be described by the following equations~\cite{Weinberg1972,Will1981,KopeikinEfroimskyKaplan2012}
\begin{eqnarray}
  &&g_{00} + 2g_{0i}\frac{dx^i}{dt} + g_{ij}\frac{dx^i}{dt}\frac{dx^j}{dt} = 0~, \label{eq:Nullcondition-a}\\
  &&\frac{d^2x^i}{dt^2} + \Gamma^i_{00} +2\Gamma^i_{0j}\frac{dx^j}{dt} +\Gamma^i_{jk}\frac{dx^j}{dt}\frac{dx^k}{dt} - \left(\Gamma^0_{00}+ 2\Gamma^0_{0j}\frac{dx^j}{dt} +\Gamma^0_{jk}\frac{dx^j}{dt}\frac{dx^k}{dt}\right)\frac{dx^i}{dt}=0~,\label{eq:geodesicEq-a}
\end{eqnarray}
{\color{black}where $\Gamma_{\alpha\beta}^\mu$ denotes the Christoffel symbol
\begin{equation*}
  \Gamma_{\alpha\beta}^\mu = \frac{1}{2}g^{\rho\mu}\left(\frac{\partial g_{\rho\beta}}{\partial x^\alpha}+\frac{\partial g_{\rho\alpha}}{\partial x^\beta} - \frac{\partial g_{\alpha\beta}}{\partial x^\rho}\right)~.
\end{equation*}}
\indent Substituting Eqs.\,\eqref{eq:metric-1nd}-\eqref{eq:metric-3nd} into Eqs.\,\eqref{eq:Nullcondition-a}-\eqref{eq:geodesicEq-a}, we can obtain the 2PN dynamics equations of photon as follows
\begin{equation}\label{eq:Nullcondition-2PN}
  -1 +\frac{2m}{r} -\frac{2m^2+q^2}{r^2} + 2\bm{\zeta}\!\cdot\!\frac{d\bm{x}}{dt} + \Big(1 +\frac{2m}{r} + \frac{m^2}{r^2}\Big)\bigg|\frac{d\bm{x}}{dt}\bigg|^2 + \frac{m^2-q^2}{r^4}\bigg(\bm{x}\!\cdot\!\frac{d\bm{x}}{dt}\bigg)^2 = 0~,\nn
\end{equation}
\begin{eqnarray}
\frac{d^2\bm{x}}{dt^2} &=& -\frac{m\bm{x}}{r^3} + \frac{(4m^2+q^2) \bm{x}}{r^4} + \frac{4m r - 2m^2-2q^2}{r^4}\bigg(\bm{x}\!\cdot\!\frac{d\bm{x}}{dt}\bigg)\frac{d\bm{x}}{dt} - \frac{(m r - q^2)\bm{x}}{r^4}\bigg|\frac{d\bm{x}}{dt}\bigg|^2 \nn\\
&&+ \frac{(2m^2-2q^2)\bm{x}}{r^6}\bigg(\bm{x}\!\cdot\!\frac{d\bm{x}}{dt}\bigg)^2 +\frac{d\bm{x}}{dt}\!\times\!(\nabla\!\times\!\bm{\zeta}) - \frac{d\bm{x}}{dt}\bigg(\frac{d\bm{x}}{dt}\!\cdot\!\bigg[\bigg(\!\frac{d\bm{x}}{dt}\!\cdot\!\nabla\!\bigg)\bm{\zeta}\bigg]\bigg)~,\label{eq:geodesicEq-2PN}\nn
\end{eqnarray}
\textcolor[rgb]{0.00,0.00,0.00}{where $\nabla$ is Nabla symbol denoting the vector differential operator~\cite{Weinberg1972}.}

Based on these two equations, we can obtain the 2PN velocity and trajectory of light propagation in the field of Kerr-Newman black hole.

\section{The 2PN solution to the light propagation}\label{sec:3rd}

Assuming a photon emitted at the coordinate time $t_{\rm e}$ at a point $\bm{x}_{\rm e}$ with an initial direction unit vector $\bm{n}$. Following Will's method~\cite{Will1981}, we can write the 2PN trajectory and velocity of light as
\begin{equation}\label{eq:trajectory-2PN-general}
  \bm{x} = \bm{x}_{\sss\rm N} + \bm{x}_{\sss\rm 1PN} + \bm{x}_{\sss\rm 2PN}~,
\end{equation}
\begin{equation}\label{eq:trajectory-2PN-general}
  \frac{d\bm{x}}{dt} = \bm{n} + \frac{d\bm{x}_{\sss\rm 1PN}}{dt} + \frac{d\bm{x}_{\sss\rm 2PN}}{dt}~,
\end{equation}
where $\bm{x}_{\sss\rm N}$ denotes the Newtonian solution of light trajectory
\begin{equation}
  \bm{x}_{\sss\rm N} = \bm{x}_{\rm e} + \bm{n}(t -t_{\rm e})~,\label{eq:trajectory-0PN}
\end{equation}
which means the light travels in a straight-line with a constant velocity of $1$. $\bm{x}_{\sss\rm 1PN}$, $\bm{x}_{\sss\rm 2PN}$ denote the 1PN and 2PN corrections to the Newtonian solution, which can be obtained via an iterative method. The derivations of these PN corrections are straightforward but tedious, thus we give them in the Appendix.
\begin{figure}[h]
  \centering
  \includegraphics[width=5in]{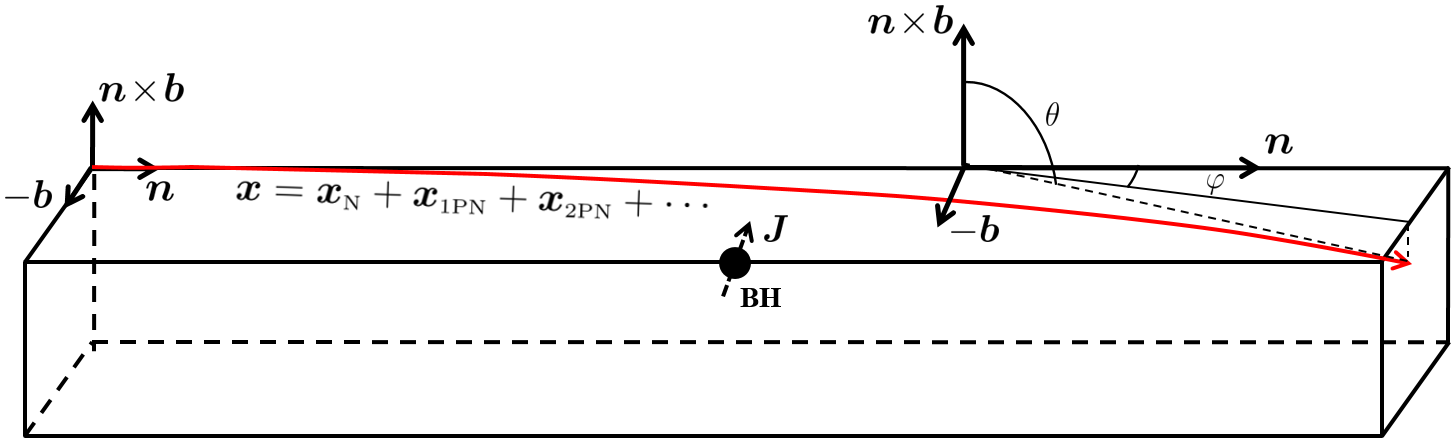}\\
  \caption{Schematic diagram for the light propagation in Kerr-Newman spacetime. The direction of the angular momentum $\bm{J}$ of black hole is arbitrary.
  The light trajectory is denoted by {\color{black}the red solid line} and in general is not in the incident plane spanned by $\bm{n}\!\times\!\bm{b}$. $\theta$ and $\varphi$ denote the light deflection angles.} \label{Fig1}
\end{figure}

Here we consider the configuration in which the Kerr-Newman black hole is at the origin, and both light source and observer lie in the asymptotically flat region, i.e., $\bm{x}_{\rm e}\!\rightarrow\!-\infty$ and $\bm{x}_{\sss\rm N}\!\rightarrow\!+\infty$. This scenario is the most interesting in astronomy observations. In this case, the 1PN and 2PN corrections to the Newtonian solution for the light trajectory can be written as follows
\begin{equation}
\bm{x}_{\sss\rm 1PN} = \bm{n}\!\bigg[\!-\!2m\ln\!{\frac{|\bm{x}_{\sss\rm N}|+\bm{n}\!\cdot\!\bm{x}_{\sss\rm N}}{|\bm{x}_{\rm e}|+\bm{n}\!\cdot\!\bm{x}_{\rm e}}}\!\bigg] +\bm{b}\bigg[\!-\!\frac{2m|\bm{x}_{\sss\rm N}|}{b^2}\!\bigg(\!1\!-\!\frac{\bm{x}_{\rm e}\!\cdot\!\bm{x}_{\sss\rm N}}{|\bm{x}_{\rm e}||\bm{x}_{\sss\rm N}|}\!\bigg)\!\bigg]~,\label{eq:trajectory-photon-1pn}
\end{equation}
\begin{eqnarray}
\hskip -0cm \bm{x}_{\sss\rm 2PN} &=& \bm{n}\bigg[\!-\!\frac{8m^2|\bm{x}_{\sss\rm N}| }{b^2} \!-\!\frac{15\pi m^2\!-\!3\pi q^2}{4b}\!-\!\frac{4(\bm{n}\!\times\!\bm{b})\!\cdot\!\bm{J}}{b^2}\bigg]\!\!+\bm{n}\!\times\!\bm{b}\bigg[\!\frac{4\bm{n}\!\cdot\!\bm{J}}{b^2} \!-\!\frac{2|\bm{x}_{\sss\rm N}|\bm{b}\!\cdot\!\bm{J}}{b^4}\!\bigg(\!1\!-\!\frac{\bm{x}_{\rm e}\!\cdot\!\bm{x}_{\sss\rm N}}{|\bm{x}_{\rm e}||\bm{x}_{\sss\rm N}|}\!\bigg)\!\bigg]\nn\\
&& +\bm{b}\bigg[\frac{8m^2}{b^2}\!\ln\!{\frac{|\bm{x}_{\sss\rm N}|\!+\!\bm{n}\!\cdot\!\bm{x}_{\sss\rm N}}{|\bm{x}_{\rm e}|\!+\!\bm{n}\!\cdot\!\bm{x}_{\rm e}}}\!-\!\frac{\!(\!15m^2\!-\!3q^2)\pi|\bm{x}_{\sss\rm N}|\!}{4b^3}\!+\!\frac{4m^2}{b^2}\!\bigg(\!1\!+\!\frac{|\bm{x}_{\sss\rm N}|}{|\bm{x}_{\rm e}|}\bigg)\!+\!\frac{m^2\!+\!3q^2}{4b^2}\!\bigg(\!1\!-\!\frac{\bm{x}_{\rm e}\!\cdot\!\bm{x}_{\sss\rm N}}{|\bm{x}_{\rm e}|^2}\!\bigg) \nn\\
&&\hskip 1cm -\frac{2|\bm{x}_{\sss\rm N}|(\bm{n}\!\times\!\bm{b})\!\cdot\!\bm{J}}{b^4}\!\bigg(\!1\!-\!\frac{\bm{x}_{\rm e}\!\cdot\!\bm{x}_{\sss\rm N}}{|\bm{x}_{\rm e}||\bm{x}_{\sss\rm N}|}\!\bigg)\bigg]~,\label{eq:trajectory-2PN}\nn
\end{eqnarray}
where $\bm{b} \equiv \bm{x}_{\sss\rm N} - \bm{n}(\bm{n}\cdot\bm{x}_{\sss\rm N})=\bm{x}_{\sss\rm e} - \bm{n}(\bm{n}\cdot\bm{x}_{\sss\rm e})$ is the impact vector joining the center of the Kerr-Newman black hole and the point of the closest approach in the line of $\bm{x}_{\sss\rm N}$, whose amplitude $b \equiv |\bm{b}|$ is well-known as the impact parameter \cite{Weinberg1972}. The plane spanned by the vectors $\bm{n}$ and $\bm{b}$ is defined as the incident plane. FIG.\,\ref{Fig1} shows the schematic diagram for the light propagation. In Eqs.\,\eqref{eq:trajectory-photon-1pn} and \eqref{eq:trajectory-2PN}, some terms other than the ones being proportional to $|\bm{x}_{\sss\rm N}|$ are kept in order to show the important PN effects.

{\color{black}The light velocity in the 2PN accuracy, given by the combination of Eqs.\,(A22) and (A23), can be written as}
\begin{equation}\label{eq:velocity-infty}
  \frac{d\bm{x}}{dt} = \bm{n}\bigg(1\! -\! \frac{8m^2}{b^2}\bigg) + \bm{b} \bigg[\!-\frac{4m}{b^2} \!-\! \frac{15\pi m^2}{4b^3} \!+\! \frac{3\pi q^2}{4b^3} \!-\! \frac{4(\bm{n}\times\bm{b})\!\cdot\!\bm{J}}{b^{4}}\bigg]
  +\bm{n}\!\times\!\bm{b}\bigg(\!-\!\frac{4\bm{b}\cdot\!\bm{J}}{b^4}\bigg)~,
\end{equation}
here we have dropped all the terms vanishing in the limits of $\bm{x}_{\sss\rm N}\!\rightarrow\! \infty$ and $\bm{x}_{\rm e}\!\rightarrow\!-\infty$.

These results can be used to calculate the light deflection caused by Kerr-Newman black hole.

\section{The 2PN light deflection in Kerr-Newman spacetime}\label{sec:4th}

The light deflection is of great importance to the phenomena of gravitational lensing.
The 2PN deflection of light in Kerr-Newman spacetime can be obtained from Eq.\,\eqref{eq:velocity-infty} and characterized by the polar angle $\theta$ and the azimuthal angle $\varphi$ in the coordinate system ($\bm{n},~\bm{b},~\bm{n}\!\times\!\bm{b}$) as follows
\begin{equation}\label{eq:deflection-infty}
  \left\{\begin{array}{c}\theta \\ \varphi\end{array}\right\} = \left\{\begin{aligned}&\hskip2cm \frac{\pi}{2}+\frac{4\bm{b}\cdot\bm{J}}{b^3}\\
  &-\frac{4m}{b} - \frac{15\pi m^2}{4b^2} + \frac{3\pi q^2}{4b^{2}} - \frac{4(\bm{n}\!\times\!\bm{b})\cdot\bm{J}}{b^3}\end{aligned}\right\}~.
\end{equation}
Notice that the azimuthal angle $\varphi$ is measured from the direction of $\bm{n}$ in the incident plane ($\theta=\pi/2$) and the polar angle $\theta$ is measured from the direction of $\bm{n}\!\times\!\bm{b}$, as shown in FIG.\ref{Fig1}. It can be observed from Eq.\,\eqref{eq:deflection-infty} that the light will deviate from the incident plane for $\bm{b}\cdot\!\bm{J}\!\neq\!0$.

\begin{figure}[th]
  \centering
  \includegraphics[width=5in]{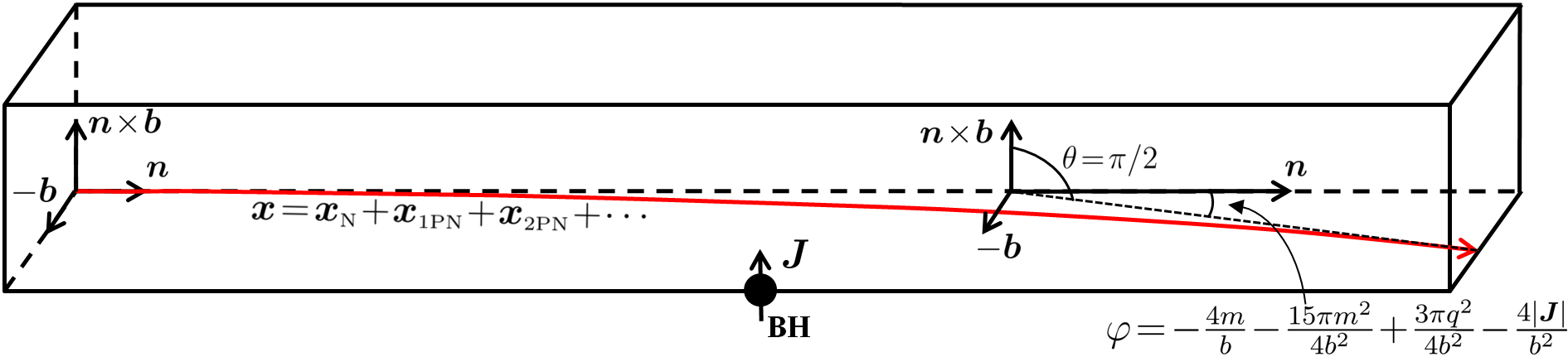}
  \caption{The schematic diagram for light propagation in the equatorial plane. The light trajectory is represented as a red {\color{black}solid} line. {\color{black}Here only the retro-orbit case is shown.}}\label{Fig2}
\end{figure}

For the light being in the equatorial plane of Kerr-Newman black hole initially, i.e., $\bm{b} \!\cdot\! \bm{J}\!=\!0$, $\bm{n} \cdot\!\bm{J}\!=\!0$, and $(\bm{n}\times\!\bm{b})\!\cdot\!\bm{J}=\pm b|\bm{J}|$, from Eq.\,\eqref{eq:deflection-infty} we have $\theta=\pi/2$ and $\varphi = -\frac{4m}{b} - \frac{15\pi m^2}{4b^2} + \frac{3\pi q^2}{4b^{\color{black}2}} {\color{black}-}\left(\pm \frac{4|\bm{J}|}{b^2}\right)$, with the sign ``$+$" corresponding to the retro-orbit case and ``$-$" corresponding to the direct-orbit one. The light will stay in the equatorial plane all the time, as shown in FIG.\ref{Fig2}. This result agrees with the equatorial deflection formula in the 2PN approximation~\cite{ChakrabortySen2015}, and reduces to the Kerr deflection formula~\cite{Edery2006} when the electric charge $q$ is set as zero.

Another case needs special attention, in which the initial velocity of light from infinity is parallel to the angular momentum, i.e., $\bm{n} \!\parallel \! \bm{J}$. In this case we have $\bm{b}\cdot\!\bm{J}\!=\!0$, $(\bm{n}\!\times\!\bm{b})\cdot\! \bm{J}=0$. From Eq.\,\eqref{eq:deflection-infty}, we have $\theta=\frac{\pi}{2}$ and {$ \varphi =-\frac{4m}{b} - \frac{15\pi m^2}{4b^2}+ \frac{3\pi q^2}{4b^{\color{black}2}}$, seeming that the rotation does not have effect. However, this is an illusion. From Eq.\,\eqref{eq:trajectory-2PN} we can see that the light trajectory has a displacement of {$\frac{4\bm{n}\cdot\bm{J}}{b}\!\left(\bm{n}\!\times\!\frac{\bm{b}}{b}\right)$} from the incident plane. In fact, this rotation effect is first found and derived by Klioner~\cite{Klioner1991}. FIG.\ref{Fig3} shows the schematic trajectory of light in this special case: the light keeps on deviating from the incident plane along with the rotation direction when it approaches to and leaves from the gravitational source, though it's final velocity will become parallel to the incident plane again when it goes to infinity.
\begin{figure}[H]
  \centering
  \includegraphics[width=5in]{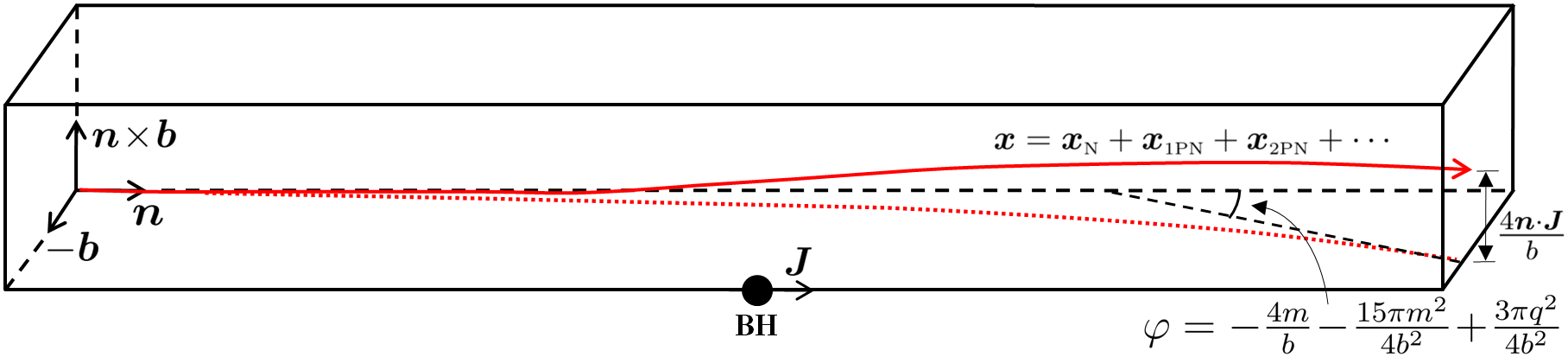}\\
  \caption{The schematic diagram for light propagation for the case of $\bm{n} \!\parallel \! \bm{J}$. The red solid line denotes the light trajectory, which deviates gradually from the incident plane but finally become parallel to the latter again in the limit of $ \bm{x}_{\sss\rm N}\!\rightarrow\! \infty $. The red dotted line denotes the projection of the light trajectory onto the incident plane from top to bottom.
  }\label{Fig3}
\end{figure}

\section{summary}\label{sec:summary}
The 2PN effects may be detected in the astrometric measurements with sub-micro-arcsecond level in angular determination in near future. In this work, we derive the 2PN solution for the light propagating in the Kerr-Newman spacetime. Especially, the expression of the 2PN solution is compact when both the light source and observer are located in the asymptotically flat regions. Based on this result, we further obtain the 2PN light deflection in the Kerr-Newman spacetime. All results are exhibited in the coordinate system constituted by the initial light-direction vector, the impact vector, and their cross-product. The direction of the incident light is arbitrary, and this makes our formalism quite general and applicable for most practical situations in gravitational lensing induced by the Kerr-Newman black hole.

\section*{ACKNOWLEDGEMENT}
\textcolor[rgb]{0.00,0.00,0.00}{We are very grateful to the referee for his/her constructive comments and suggestions for improving the quality of this paper.} This work was supported in part by the National Natural Science Foundation of China (Grant Nos. 11647161, 11647314, 11547311).

\appendix

\section{Derivations of the 2PN solution to the light propagation in Kerr-Newman spacetime}

We first recall Will's approach~\cite{Will1981} to get the 1PN equations of the photon motion.

Considering a photon emitted at the coordinate time $t_{\rm e}$ at a point $\bm{x}_{\rm e}$ in an initial direction described by the unit vector. For the zeroth order, Eqs.\,\eqref{eq:Nullcondition-2PN} and \eqref{eq:geodesicEq-2PN} reduce to
\begin{eqnarray}
\bigg|\frac{d\bm{x}}{dt}\bigg|= 1~,~~~~\frac{d^2\bm{x}}{dt^2} =0~,\label{eq:NewtonianEquation}
\end{eqnarray}
and the zeroth-order solution (Newtonian solution) is
\begin{equation}
  \frac{d\bm{x}_{\sss\rm N}}{dt} = \bm{n}~,\label{NewtonianMotionV}
\end{equation}
\begin{equation}
  \bm{x}_{\sss\rm N} = \bm{x}_{\rm e} + \bm{n}(t -t_{\rm e})~.\label{NewtonianMotionT}
\end{equation}

To 1PN accuracy, Eqs.\,\eqref{eq:Nullcondition-2PN} and \eqref{eq:geodesicEq-2PN} reduce to
\begin{equation}
-1 +\frac{2m}{r} + \Big(1 +\frac{2m}{r}\Big)\bigg|\frac{d\bm{x}}{dt}\bigg|^2 = 0~,\label{eq:Nullcondition-1PN}
\end{equation}
\begin{equation}
\frac{d^2\bm{x}}{dt^2} = -\bigg(1 + \bigg|\frac{d\bm{x}}{dt}\bigg|^2\bigg)\frac{m x^i}{r^3} + \frac{4m}{r^3}\frac{dx^i}{dt}\bigg(\bm{x}\cdot\frac{d\bm{x}}{dt}\bigg)~,\label{eq:geodesicEq-1PN}
\end{equation}
and the corresponding solution has the form
\begin{equation}
  \bm{x} = \bm{x}_{\sss\rm N} + \bm{x}_{\sss\rm 1PN}~,\label{eq:trajectory-1PN-form}
\end{equation}
with $\bm{x}_{\sss\rm 1PN}$ being the 1PN correction.

Substituting Eqs.\,\eqref{NewtonianMotionT} and \eqref{eq:trajectory-1PN-form} into Eqs.\,\eqref{eq:Nullcondition-1PN}-\eqref{eq:geodesicEq-1PN}, and only keeping the 1PN terms, we can obtain
\begin{equation}\label{eq:v-velocity-1PN_dynamics}
  \bm{n}\cdot\frac{d\bm{x}_{\sss\rm 1PN}}{dt} = -\frac{2m}{|\bm{x}_{\sss\rm N}|}~.
\end{equation}
\begin{equation}\label{eq:acceleration-1PN_dynamics}
  \frac{d^2\bm{x}_{\sss\rm 1PN}}{dt^2} = -\frac{2m}{|\bm{x}_{\sss\rm N}|^3}\bm{x}_{\sss\rm N} + \frac{4m}{|\bm{x}_{\sss\rm N}|^3}\bm{n}(\bm{n}\cdot\bm{x}_{\sss\rm N})~.
\end{equation}

In order to solve these two equations, we decompose $\bm{x}_{\sss\rm 1PN}$ into components parallel and perpendicular to $\bm{n}$:
\begin{eqnarray}
  && \bm{x}_{\sss\rm 1PN\parallel} = \bm{n}(\bm{n}\cdot\bm{x}_{\sss\rm 1PN})~, \label{eq:v-velocity-1PN_Parallel}\\
  && \bm{x}_{\sss\rm 1PN\perp} = \bm{x}_{\sss\rm 1PN} - \bm{n}(\bm{n}\cdot\bm{x}_{\sss\rm 1PN})~.\label{eq:acceleration-1PN_Perp}
\end{eqnarray}
Eqs.\,\eqref{eq:v-velocity-1PN_dynamics} and \eqref{eq:acceleration-1PN_dynamics} then yield
\begin{eqnarray}
  \frac{d\bm{x}_{\sss\rm 1PN\parallel}}{dt} &=& -\frac{2m\bm{n}}{|\bm{x}_{\sss\rm N}|}~,\label{eq:velocity-parallel-1PN}\\
  \frac{d^2\bm{x}_{\sss\rm 1PN\perp}}{dt^2} &=& -\frac{2m}{|\bm{x}_{\sss\rm N}|^3}\bm{x}_{\sss\rm N} + \frac{2m}{|\bm{x}_{\sss\rm N}|^3}\bm{n}(\bm{n}\cdot\bm{x}_{\sss\rm N}) = -\frac{2m\bm{b}}{|\bm{x}_{\sss\rm N}|^3}~{\color{black}.}\label{eq:acceleration-perp-1PN}
\end{eqnarray}

Integrating Eq.\,\eqref{eq:acceleration-perp-1PN} along $\bm{x}_{\sss\rm N}$, we get
\begin{equation}\label{eq:velocity-1PN-perp}
  \frac{d\bm{x}_{\sss\rm 1PN\perp}}{dt} = - \frac{2m\bm{b}}{b^2}\bigg(\frac{\bm{n}\cdot\bm{x}_{\sss\rm N}}{|\bm{x}_{\sss\rm N}|} - \frac{\bm{n}\cdot\bm{x}_{\rm e}}{|\bm{x}_{\rm e}|}\bigg)~.
\end{equation}
Combining Eqs.\,\eqref{eq:velocity-parallel-1PN} and \eqref{eq:velocity-1PN-perp}, we have the velocity of the photon to the 1PN accuracy
\begin{equation}\label{eq:velocity-photon-1PN}
  \frac{d\bm{x}_{\sss\rm 1PN}}{dt} = -\frac{2m\bm{n}}{|\bm{x}_{\sss\rm N}|} - \frac{2m\bm{b}}{b^2}\bigg(\frac{\bm{n}\cdot\bm{x}_{\sss\rm N}}{|\bm{x}_{\sss\rm N}|} - \frac{\bm{n}\cdot\bm{x}_{\rm e}}{|\bm{x}_{\rm e}|}\bigg)~.
\end{equation}
Notice that this equation can also be directly retrieved from Will's results based on the parameterized post-Newtonian approximation~\cite{Will1981}, but for completeness of this work, we include this part here.

Integrating Eq.\,\eqref{eq:velocity-photon-1PN} along the straight-line $\bm{x}_{\sss\rm N}$, we can obtain the 1PN correction to the trajectory of light as follow
\begin{equation}\label{eq:trajectory-photon-1pn_Appendix}
  \bm{x}_{\sss\rm 1PN} = -2m\bm{n}\ln{\frac{|\bm{x}_{\sss\rm N}|+\bm{n}\cdot\bm{x}_{\sss\rm N}}{|\bm{x}_{\rm e}|+\bm{n}\cdot\bm{x}_{\rm e}}} - \frac{2m|\bm{x}_{\sss\rm N}|\bm{b}}{b^2}\bigg(1 - \frac{\bm{x}_{\rm e}\cdot\bm{x}_{\sss\rm N}}{|\bm{x}_{e}||\bm{x}_{\sss\rm N}|}\bigg).
\end{equation}
Following the same procedure, we can further deduce the 2PN equations of light propagation using the iterative method.

To 2PN accuracy, the solution of Eqs.\,\eqref{eq:Nullcondition-2PN} and \eqref{eq:geodesicEq-2PN} can be written as
\begin{equation}\label{eq:trajectory-2PN-general}
  \bm{x} = \bm{x}_{\sss\rm N} + \bm{x}_{\sss\rm 1PN}+ \bm{x}_{\sss\rm 2PN}~,
\end{equation}
with $\bm{x}_{\sss\rm 2PN}$ being the 2PN correction.

Plugging Eq.\,\eqref{eq:trajectory-2PN-general} into Eq.\,\eqref{eq:Nullcondition-2PN}, and making use of Eqs.\,\eqref{NewtonianMotionT}, \eqref{eq:velocity-photon-1PN}-\eqref{eq:trajectory-photon-1pn_Appendix}, we obtain
\begin{eqnarray}
\bm{n}\!\cdot\!\frac{d\bm{x}_{\sss\rm2PN}}{dt} &=& - \frac{4m^2(\bm{n}\!\cdot\!\bm{x}_{\sss\rm N})}{|\bm{x}_{\sss\rm N}|^3}\!\ln\!{\frac{|\bm{x}_{\sss\rm N}|\!+\!\bm{n}\!\cdot\!\bm{x}_{\sss\rm N}}{|\bm{x}_{\rm e}|\!+\!\bm{n}\!\cdot\!\bm{x}_{\rm e}}}\!+\!\frac{q^2}{|\bm{x}_{\sss\rm N}|^2} \!-\!\frac{4m^2}{|\bm{x}_{\rm e}||\bm{x}_{\sss\rm N}|} \!+\!\frac{2m^2}{|\bm{x}_{\rm e}|^2}\!-\!\frac{4m^2}{b^2} \nn\\
&& +\bigg(\frac{4m^2}{|\bm{x}_{\sss\rm N}|^2}\!+\!\frac{4m^2}{b^2}\bigg)\!\frac{\bm{x}_{\rm e}\!\cdot\bm{x}_{\sss\rm N}}{|\bm{x}_{\rm e}||\bm{x}_{\sss\rm N}|}\!+\!\frac{(m^2\!-\!q^2)b^2}{2|\bm{x}_{\sss\rm N}|^4} -\bm{n}\cdot\bm{\zeta}~, \label{eq:v-velocity-2PN}
\end{eqnarray}
here and from now on the vector potential $\bm{\zeta}$ is evaluated by $2(\bm{x}_{\sss\rm N}\!\times\!\bm{J})/|\bm{x}_{\sss\rm N}|^3$.

Substituting Eq.\,\eqref{eq:trajectory-2PN-general} into Eq.\,\eqref{eq:geodesicEq-2PN}, making use of Eqs.\,\eqref{NewtonianMotionT},\,\eqref{eq:acceleration-1PN_dynamics},\,\eqref{eq:velocity-photon-1PN}-\eqref{eq:trajectory-photon-1pn_Appendix}, we can obtain
\begin{eqnarray}
\frac{d^2\bm{x}_{\sss\rm2PN}}{dt^2} &=& \bm{n}\bigg\{\frac{2m^2}{|\bm{x}_{\sss\rm N}|^3}\!\bigg[\bigg(\!4\!-\!\frac{6b^2}{|\bm{x}_{\sss\rm N}|^2}\!\bigg)\!\ln\!{\frac{|\bm{x}_{\sss\rm N}|\!+\!\bm{n}\!\cdot\!\bm{x}_{\sss\rm N}}{|\bm{x}_{\rm e}|\!+\!\bm{n}\!\cdot\!\bm{x}_{\rm e}}}\!-\!\frac{2\bm{n}\!\cdot\!\bm{x}_{\rm e}}{|\bm{x}_{\rm e}|} \!-\!\frac{2\bm{n}\!\cdot\!\bm{x}_{\sss\rm N}}{|\bm{x}_{\sss\rm N}|} \!-\!\frac{b^2(\bm{n}\!\cdot\!\bm{x}_{\sss\rm N})}{|\bm{x}_{\sss\rm N}|^3}\!-\!\frac{6b^2({\color{black}\!\bm{n}\!\cdot\!\bm{x}_{\sss\rm N}\!-\!\bm{n}\!\cdot\!\bm{x}_{\rm e}\!})}{|\bm{x}_{\rm e}||\bm{x}_{\sss\rm N}|^2}\bigg] \!\nn\\
&&- \frac{2q^2(\bm{n}\!\cdot\!\bm{x}_{\sss\rm N})^3}{|\bm{x}_{\sss\rm N}|^6}\!-\!\bm{n}\!\cdot\![(\bm{n}\!\cdot\!\nabla)\bm{\zeta}]\bigg\}+\bm{b}\bigg\{\frac{2m^2}{|\bm{x}_{\sss\rm N}|^3}\!\bigg[\!-\!\frac{6\bm{n}\!\cdot\!\bm{x}_{\sss\rm N}}{|\bm{x}_{\sss\rm N}|^2}\!\ln\!{\frac{|\bm{x}_{\sss\rm N}|\!+\!\bm{n}\!\cdot\!\bm{x}_{\sss\rm N}}{|\bm{x}_{\rm e}|\!+\!\bm{n}\!\cdot\!\bm{x}_{\rm e}}}\!+\! \frac{3}{|\bm{x}_{\sss\rm N}|}\!-\!\frac{4}{|\bm{x}_{\rm e}|} \nn\\
&&-\frac{2|\bm{x}_{\sss\rm N}|}{b^2} \!-\!\frac{b^2}{|\bm{x}_{\sss\rm N}|^3}\!+\! \bigg(\!\frac{6}{|\bm{x}_{\sss\rm N}|^2}\!+\!\frac{2}{b^2}\!\bigg)\!\frac{\bm{x}_{\rm e}\!\cdot\!\bm{x}_{\sss\rm N}}{|\bm{x}_{\rm e}|}\bigg]+\frac{2q^2b^2}{|\bm{x}_{\sss\rm N}|^6}\bigg\}+\bm{n}\!\times\!(\nabla\!\times\!\bm{\zeta})~.
\label{eq:acceleration-2PN_dynamics}
\end{eqnarray}

Similarly, we can decompose $\bm{x}_{\sss\rm 2PN}$ into components parallel and perpendicular to $\bm{n}$:
\begin{align}
  \bm{x}_{\sss\rm 2PN\parallel} & = \bm{n}(\bm{n}\cdot\bm{x}_{\sss\rm 2PN})~, \label{2PNparallel} \\
  \bm{x}_{\sss\rm 2PN\perp} & = \bm{x}_{\sss\rm 2PN} - \bm{n}(\bm{n}\cdot\bm{x}_{\sss\rm 2PN})~. \label{2PNperp}
\end{align}
From Eqs. \eqref{eq:acceleration-2PN_dynamics} and \eqref{2PNperp} we have
\begin{eqnarray}
\frac{d^2\bm{x}_{\sss\rm2PN\perp}}{dt^2} &=& \frac{2m^2 \bm{b}}{|\bm{x}_{\sss\rm N}|^3}\!\bigg[\! \bigg(\!\frac{6}{|\bm{x}_{\sss\rm N}|} \!+\! \frac{2|\bm{x}_{\sss\rm N}|}{b^2}\!\bigg)\! \frac{\bm{x}_{\rm e}\!\cdot\!\bm{x}_{\sss\rm N}}{|\bm{x}_{\rm e}||\bm{x}_{\sss\rm N}|}\!-\!\frac{6\bm{n}\!\cdot\!\bm{x}_{\sss\rm N}}{|\bm{x}_{\sss\rm N}|^2}\!\ln\!{\frac{|\bm{x}_{\sss\rm N}| \!+\! \bm{n}\!\cdot\!\bm{x}_{\sss\rm N}}{|\bm{x}_{\rm e}| \!+\! \bm{n}\!\cdot\!\bm{x}_{\rm e}}} \!+\! \frac{3}{|\bm{x}_{\sss\rm N}|} \!-\! \frac{4}{|\bm{x}_{\rm e}|} \!-\! \frac{2|\bm{x}_{\sss\rm N}|}{b^2} \!-\! \frac{b^2}{|\bm{x}_{\sss\rm N}|^3}\!\bigg] \nn\\
&&+\frac{2q^2 b^2\bm{b}}{|\bm{x}_{\sss\rm N}|^6} +\bm{n}\!\times\!(\nabla\!\times\!\bm{\zeta})~.\label{eq:acceleration-perp-2PN}
\end{eqnarray}
Integrating Eq.\,\eqref{eq:acceleration-perp-2PN} along $\bm{x}_{\sss\rm N}$, we can obtain
\begin{eqnarray}
\frac{d\bm{x}_{\sss\rm2PN\perp}}{dt} &=& \!\bm{b}\bigg\{\!\frac{4m^2\!}{|\bm{x}_{\sss\rm N}|^3}\!\ln\!{\frac{\!|\bm{x}_{\sss\rm N}|\!+\!\bm{n}\!\cdot\!\bm{x}_{\sss\rm N}\!}{|\bm{x}_{\rm e}| \!+\! \bm{n}\!\cdot\!\bm{x}_{\rm e}}} \!+\! \frac{\!15m^2\!-\!3q^2\!}{4b^3}\!\bigg(\!\!\arccos\!{\frac{\!\bm{n}\!\cdot\!\bm{x}_{\sss\rm N}\!}{|\bm{x}_{\sss\rm N}|}} \!-\! \arccos\!{\frac{\!\bm{n}\!\cdot\!\bm{x}_{\rm e}\!}{|\bm{x}_{\rm e}|}}\!\bigg) \!+\!\frac{4m^2(b^2\!+\!|\bm{x}_{\sss\rm N}|^2)({\color{black}\!\bm{n}\!\cdot\!\bm{x}_{\sss\rm N}\!-\!\bm{n}\!\cdot\!\bm{x}_{\rm e}\!})}{b^2|\bm{x}_{\sss\rm N}|^3|\bm{x}_{\rm e}|}\nn\\
&& \hskip 0.25cm +\frac{m^2\!+\!3q^2}{4b^2}\!\bigg(\!\frac{\bm{n}\!\cdot\!\bm{x}_{\sss\rm N}}{|\bm{x}_{\sss\rm N}|^2} \!-\!\frac{\bm{n}\!\cdot\!\bm{x}_{\rm e}}{|\bm{x}_{\rm e}|^2}\!\bigg) \!-\! \frac{\!m^2\!-\!q^2\!}{2}\!\bigg(\!\frac{\bm{n}\!\cdot\!\bm{x}_{\sss\rm N}}{|\bm{x}_{\sss\rm N}|^4} \!-\! \frac{\bm{n}\!\cdot\!\bm{x}_{\rm e}}{|\bm{x}_{\rm e}|^4}\!\bigg) \!-\!\frac{2(\bm{n}\!\times\!\bm{b})\!\cdot\!\bm{J}}{b^4}\!\bigg(\!\frac{\bm{n}\!\cdot\!\bm{x}_{\sss\rm N}}{|\bm{x}_{\sss\rm N}|} \!-\!\frac{\bm{n}\!\cdot\!\bm{x}_{\rm e}}{|\bm{x}_{\rm e}|}\!\bigg)\!\bigg\} \nn\\
&& +\bm{n}\!\times\!\bm{b}\bigg\{\!2\bm{n}\!\cdot\!\bm{J}\bigg(\!\frac{1}{|\bm{x}_{\sss\rm N}|^3} \!-\!\frac{1}{|\bm{x}_{\rm e}|^3}\!\bigg) \!-\! \frac{2\bm{b}\!\cdot\!\bm{J}}{b^2}\!\bigg(\!\frac{\bm{n}\!\cdot\!\bm{x}_{\sss\rm N}}{b^2|\bm{x}_{\sss\rm N}|} \!-\!\frac{\bm{n}\!\cdot\!\bm{x}_{\rm e}}{b^2|\bm{x}_{\rm e}|}\!+\!\frac{\bm{n}\!\cdot\!\bm{x}_{\sss\rm N}}{|\bm{x}_{\sss\rm N}|^3} \!-\!\frac{\bm{n}\!\cdot\!\bm{x}_{\rm e}}{|\bm{x}_{\rm e}|^3}\!\bigg)\!\bigg\}~. \label{eq:velocity-2PN-perp}
\end{eqnarray}
From Eqs. \eqref{eq:v-velocity-2PN} and \eqref{2PNparallel} we have
\begin{eqnarray}
\frac{d\bm{x}_{\sss\rm 2PN\parallel}}{dt} &=& \bm{n}\bigg\{\!\!-\!\frac{4m^2(\bm{n}\!\cdot\!\bm{x}_{\sss\rm N})}{|\bm{x}_{\sss\rm N}|^3}\!\ln\!{\frac{|\bm{x}_{\sss\rm N}|+\bm{n}\!\cdot\!\bm{x}_{\sss\rm N}}{|\bm{x}_{\rm e}|+\bm{n}\!\cdot\!\bm{x}_{\rm e}}}\!-\!\frac{4m^2}{b^2} \!+\! \frac{2m^2}{|\bm{x}_{\rm e}|^2} \!+\!\frac{m^2b^2}{2|\bm{x}_{\sss\rm N}|^4} \!-\!\frac{4m^2}{|\bm{x}_{\rm e}||\bm{x}_{\sss\rm N}|} \nn\\
&& \hskip.5cm+ \bigg({\color{black}\frac{4m^2}{|\bm{x}_{\sss\rm N}|^2}} \!+\! \frac{4m^2}{b^2}\bigg)\!\frac{\bm{x}_{\rm e}\!\cdot\!\bm{x}_{\sss\rm N}}{|\bm{x}_{\rm e}||\bm{x}_{\sss\rm N}|} \!+\!\frac{q^2}{|\bm{x}_{\sss\rm N}|^2} \!-\!\frac{q^2b^2}{2|\bm{x}_{\sss\rm N}|^4} - \bm{n}\cdot\bm{\zeta}\bigg\}~{\color{black}.}\label{eq:velocity-parallel-2PN}
\end{eqnarray}
The summation of Eqs.\,\eqref{eq:velocity-2PN-perp} and \eqref{eq:velocity-parallel-2PN} constitutes the 2PN correction to the light velocity $\frac{d\bm{x}_{\sss\rm2PN}}{dt}$, and the 2PN correction to the trajectory $\bm{x}_{\sss\rm 2PN}$ can be achieved via integrating it along the straight-line $\bm{x}_{\sss\rm N}$ as follow
\begin{eqnarray}
  \bm{x}_{\sss\rm 2PN} \!&=& \!\bm{n}\bigg\{\!\frac{4m^2\!}{|\bm{x}_{\sss\rm N}|}\!\ln\!{\frac{|\bm{x}_{\sss\rm N}|\!+\!\bm{n}\!\cdot\!\bm{x}_{\sss\rm N}}{|\bm{x}_{\rm e}|\!+\!\bm{n}\!\cdot\!\bm{x}_{\rm e}}} \!+\!\frac{\!15m^2\!-\!3q^2\!}{4b}\!\bigg(\!\!\arccos\!{\frac{\bm{n}\!\cdot\!\bm{x}_{\sss\rm N}\!}{|\bm{x}_{\sss\rm N}|}}\!-\!\arccos\!{\frac{\bm{n}\!\cdot\!\bm{x}_{\rm e}\!}{|\bm{x}_{\rm e}|}}\!\bigg) \!+\!\frac{\!m^2\!-\!q^2\!}{4}\!\bigg(\!\frac{\bm{n}\!\cdot\!\bm{x}_{\sss\rm N}}{|\bm{x}_{\sss\rm N}|^2} \!-\!\frac{\bm{n}\!\cdot\!\bm{x}_{\rm e}}{|\bm{x}_{\rm e}|^2}\!\bigg) \nn\\
  && \hskip .5cm+\frac{2m^2}{|\bm{x}_{\rm e}|}\!\bigg(\!\frac{2}{|\bm{x}_{\sss\rm N}|} \!+\!\frac{1}{|\bm{x}_{\rm e}|}\!\bigg)\!(\!\bm{n}\!\cdot\!\bm{x}_{\sss\rm N}\!-\!\bm{n}\!\cdot\!\bm{x}_{\rm e}\!)\!-\!\frac{4m^2 |\bm{x}_{\sss\rm N}|\!+\!2(\bm{n}\!\times\!\bm{b})\!\cdot\!\bm{J}}{b^2}\!\bigg(\!\frac{\bm{n}\!\cdot\!\bm{x}_{\sss\rm N}}{|\bm{x}_{\sss\rm N}|} \!-\! \frac{\bm{n}\!\cdot\!\bm{x}_{\rm e}}{|\bm{x}_{\rm e}|}\!\bigg)\!\bigg\} \nn\\
  &&\hskip -1.25cm +\bm{n}\!\times\!\bm{b}\bigg\{\!\frac{2\bm{n}\!\cdot\!\bm{J}}{b^2}\!\bigg[\!\frac{\bm{n}\!\cdot\!\bm{x}_{\sss\rm N}}{|\bm{x}_{\sss\rm N}|}\!-\!\frac{\!\bm{n}\!\cdot\!\bm{x}_{\rm e}}{|\bm{x}_{\rm e}|}\!-\!\frac{b^2(\!\bm{n}\!\cdot\!\bm{x}_{\sss\rm N}\!-\!\bm{n}\!\cdot\!\bm{x}_{\rm e}\!)}{|\bm{x}_{\rm e}|^3}\!\bigg]\!\!+\!\frac{2\bm{b}\!\cdot\!\bm{J}}{b^2}\!\bigg[\!\frac{1}{|\bm{x}_{\sss\rm N}|}\!-\!\frac{2}{\!|\bm{x}_{\rm e}|\!}\!-\!\frac{\!|\bm{x}_{\sss\rm N}|\!}{b^2}\!+\!\frac{\bm{x}_{\rm e}\!\cdot\!\bm{x}_{\sss\rm N}\!}{|\bm{x}_{\rm e}|}\!\bigg(\!\frac{1}{b^2}\!+\!\frac{1}{|\bm{x}_{\rm e}|^2}\!\bigg)\!\bigg]\!\bigg\} \nn\\
  &&\hskip -0.25cm +\bm{b}\bigg\{\!\frac{4m^2}{b^2}\!\bigg(\!\frac{\bm{n}\!\cdot\!\bm{x}_{\sss\rm N}}{|\bm{x}_{\sss\rm N}|}\!-\!\frac{\bm{n}\!\cdot\!\bm{x}_{\rm e}}{|\bm{x}_{\rm e}|}\!\bigg)\!\ln\!{\frac{|\bm{x}_{\sss\rm N}|\!+\!\bm{n}\!\cdot\!\bm{x}_{\sss\rm N}}{|\bm{x}_{\rm e}|\!+\!\bm{n}\!\cdot\!\bm{x}_{\rm e}}} \!+\!\frac{\bm{n}\!\cdot\!\bm{x}_{\sss\rm N}(15m^2\!-\!3q^2)}{4b^3}\!\bigg(\!\!\arccos\!{\frac{\bm{n}\!\cdot\!\bm{x}_{\sss\rm N}}{|\bm{x}_{\sss\rm N}|}}\!-\!\arccos\!{\frac{\bm{n}\!\cdot\!\bm{x}_{\rm e}}{|\bm{x}_{\rm e}|}}\!\!\bigg) \nn\\
  &&\hskip .5cm +\frac{m^2\!-\!q^2}{4}\!\bigg(\!\frac{1}{|\bm{x}_{\sss\rm N}|^2}\!-\!\frac{1}{|\bm{x}_{\rm e}|^2}\!\!\bigg)\!+\!\frac{4m^2}{b^2}\!\bigg(\!\frac{|\bm{x}_{\sss\rm N}|}{|\bm{x}_{\rm e}|}\!-\!\frac{\bm{x}_{\rm e}\!\cdot\!\bm{x}_{\sss\rm N}}{|\bm{x}_{\rm e}||\bm{x}_{\sss\rm N}|}\!\bigg)\!+\!\bigg(\!\frac{m^2\!+\!3q^2}{4b^2}\!-\!\frac{m^2\!-\!q^2}{2|\bm{x}_{\rm e}|^2}\!\bigg)\!\bigg(\!1\!-\!\frac{\!\bm{x}_{\rm e}\!\cdot\!\bm{x}_{\sss\rm N}}{|\bm{x}_{\rm e}|^2}\!\!\bigg) \nn\\
  &&\hskip .5cm- \frac{2|\bm{x}_{\sss\rm N}|(\bm{n}\!\times\!\bm{b})\!\cdot\!\bm{J}}{b^4}\!\bigg(\!1\!-\!\frac{\bm{x}_{\rm e}\!\cdot\!\bm{x}_{\sss\rm N}}{|\bm{x}_{\rm e}||\bm{x}_{\sss\rm N}|}\!\bigg)\!\bigg\}~.\nn
  \label{eq:trajectory-2PN_Appendix}
\end{eqnarray}

In principle, with the exact harmonic metric of Kerr-Newman black hole~\cite{LinJiang2014}, the higher-order PN velocity and trajectory of light in Kerr-Newman spacetime can be achieved via the same procedure. The integrals we used in this section are listed in the Appendix~\ref{Integrals} for readers' convenience.

\section{Lists of integrals}\label{Integrals}
\vskip -0.85cm
\begin{eqnarray*}
&&\int_{t_{\rm e}}^t \!dt \!=\! \bm{n}\!\cdot\!\bm{x}_{\sss\rm N} \!-\! \bm{n}\!\cdot\!\bm{x}_{\rm e}, \\
&&\int_{t_{\rm e}}^t \!\frac{1}{|\bm{x}_{\sss\rm N}|} dt \!=\! \ln\!{\frac{|\bm{x}_{\sss\rm N}|\!+\!\bm{n}\!\cdot\!\bm{x}_{\sss\rm N}}{|\bm{x}_{\rm e}|\!+\!\bm{n}\!\cdot\!\bm{x}_{\rm e}}},\\
&&\int_{t_{\rm e}}^t \!\frac{\bm{n}\!\cdot\!\bm{x}_{\sss\rm N}}{|\bm{x}_{\sss\rm N}|} dt \!=\! |\bm{x}_{\sss\rm N}| \!-\! |\bm{x}_{\rm e}|, \\
&&\int_{t_{\rm e}}^t \!\frac{1}{|\bm{x}_{\sss\rm N}|^2} dt \!=\! -\frac{1}{b}\!\bigg(\!\arccos\!{\frac{\bm{n}\!\cdot\!\bm{x}_{\sss\rm N}}{|\bm{x}_{\sss\rm N}|}} \!-\! \arccos\!{\frac{\bm{n}\!\cdot\!\bm{x}_{\rm e}}{|\bm{x}_{\rm e}|}}\!\bigg), \\
&&\int_{t_{\rm e}}^t \!\frac{\bm{n}\!\cdot\!\bm{x}_{\sss\rm N}}{|\bm{x}_{\sss\rm N}|^2} dt \!=\! \ln\!{\frac{|\bm{x}_{\sss\rm N}|}{|\bm{x}_{\rm e}|}}, \\
&&\int_{t_{\rm e}}^t \!\frac{1}{|\bm{x}_{\sss\rm N}|^3} dt \!=\! \frac{1}{b^2}\!\bigg(\!\frac{\bm{n}\!\cdot\!\bm{x}_{\sss\rm N}}{|\bm{x}_{\sss\rm N}|} \!-\! \frac{\bm{n}\!\cdot\!\bm{x}_{\rm e}}{|\bm{x}_{\rm e}|}\!\bigg), \\
&&\int_{t_{\rm e}}^t \!\frac{\bm{n}\!\cdot\!\bm{x}_{\sss\rm N}}{|\bm{x}_{\sss\rm N}|^3} dt \!=\! -\frac{1}{|\bm{x}_{\sss\rm N}|} \!+\! \frac{1}{|\bm{x}_{\rm e}|}, \\
&&\int_{t_{\rm e}}^t \!\frac{1}{|\bm{x}_{\sss\rm N}|^4} dt \!=\! -\frac{1}{2b^3}\!\bigg(\!\arccos\!{\frac{\bm{n}\!\cdot\!\bm{x}_{\sss\rm N}}{|\bm{x}_{\sss\rm N}|}} \!-\! \arccos\!{\frac{\bm{n}\!\cdot\!\bm{x}_{\rm e}}{|\bm{x}_{\rm e}|}}\!\bigg) \!+\! \frac{1}{2b^2}\!\bigg(\!\frac{\bm{n}\!\cdot\!\bm{x}_{\sss\rm N}}{|\bm{x}_{\sss\rm N}|^2} \!-\! \frac{\bm{n}\!\cdot\!\bm{x}_{\rm e}}{|\bm{x}_{\rm e}|^2}\!\bigg), \\
&&\int_{t_{\rm e}}^t \!\frac{\bm{n}\!\cdot\!\bm{x}_{\sss\rm N}}{|\bm{x}_{\sss\rm N}|^4} dt \!=\! -\frac{1}{2|\bm{x}_{\sss\rm N}|^2} \!+\! \frac{1}{2|\bm{x}_{\rm e}|^2}, \\
&&\int_{t_{\rm e}}^t \!\frac{1}{|\bm{x}_{\sss\rm N}|^5} dt \!=\! \frac{1}{3b^2}\!\bigg(\!\frac{\bm{n}\!\cdot\!\bm{x}_{\sss\rm N}}{|\bm{x}_{\sss\rm N}|^3} \!-\! \frac{\bm{n}\!\cdot\!\bm{x}_{\rm e}}{|\bm{x}_{\rm e}|^3}\!\bigg) \!+\! \frac{2}{3b^4}\!\bigg(\!\frac{\bm{n}\!\cdot\!\bm{x}_{\sss\rm N}}{|\bm{x}_{\sss\rm N}|} \!-\! \frac{\bm{n}\!\cdot\!\bm{x}_{\rm e}}{|\bm{x}_{\rm e}|}\!\bigg), \\
&&\int_{t_{\rm e}}^t \!\frac{\bm{n}\!\cdot\!\bm{x}_{\sss\rm N}}{|\bm{x}_{\sss\rm N}|^5} dt \!=\! -\frac{1}{3|\bm{x}_{\sss\rm N}|^3} \!+\! \frac{1}{3|\bm{x}_{\rm e}|^3}, \\
&&\int_{t_{\rm e}}^t \!\frac{1}{|\bm{x}_{\sss\rm N}|^6} dt \!=\! -\frac{3}{8b^5}\!\bigg(\!\arccos\!{\frac{\bm{n}\!\cdot\!\bm{x}_{\sss\rm N}}{|\bm{x}_{\sss\rm N}|}} \!-\! \arccos\!{\frac{\bm{n}\!\cdot\!\bm{x}_{\rm e}}{|\bm{x}_{\rm e}|}}\!\bigg) \!+\! \frac{3}{8b^4}\!\bigg(\!\frac{\bm{n}\!\cdot\!\bm{x}_{\sss\rm N}}{|\bm{x}_{\sss\rm N}|^2} \!-\! \frac{\bm{n}\!\cdot\!\bm{x}_{\rm e}}{|\bm{x}_{\rm e}|^2}\!\bigg) \!+\! \frac{1}{4b^2}\!\bigg(\!\frac{\bm{n}\!\cdot\!\bm{x}_{\sss\rm N}}{|\bm{x}_{\sss\rm N}|^4} \!-\! \frac{\bm{n}\!\cdot\!\bm{x}_{\rm e}}{|\bm{x}_{\rm e}|^4}\!\bigg), \\
&&\int_{t_{\rm e}}^t \!\frac{\bm{n}\!\cdot\!\bm{x}_{\sss\rm N}}{|\bm{x}_{\sss\rm N}|^6} dt \!=\! -\frac{1}{4|\bm{x}_{\sss\rm N}|^4} \!+\! \frac{1}{4|\bm{x}_{\rm e}|^4}, \\
&&\int_{t_{\rm e}}^t \!\arccos\!{\frac{\bm{n}\!\cdot\!\bm{x}_{\sss\rm N}}{|\bm{x}_
{\sss\rm N}|}} dt \!=\! (\bm{n}\!\cdot\!\bm{x}_{\sss\rm N})\!\arccos\!{\frac{\bm{n}\!\cdot\!\bm{x}_{\sss\rm N}}{|\bm{x}_{\sss\rm N}|}} \!-\! (\bm{n}\!\cdot\!\bm{x}_{\rm e})\!\arccos\!{\frac{\bm{n}\!\cdot\!\bm{x}_{\rm e}}{|\bm{x}_{\rm e}|}} \!+\! b \ln\!{\frac{|\bm{x}_{\sss\rm N}|}{|\bm{x}_{\rm e}|}},\\
&&\int_{t_{\rm e}}^t \!\frac{1}{|\bm{x}_{\sss\rm N}|^3} \!\ln\!{\frac{|\bm{x}_{\sss\rm N}|\!+\!\bm{n}\!\cdot\!\bm{x}_{\sss\rm N}}{|\bm{x}_{\rm e}|\!+\!\bm{n}\!\cdot\!\bm{x}_{\rm e}}} dt\!=\! \frac{1}{b^2}\frac{\bm{n}\!\cdot\!\bm{x}_{\sss\rm N}}{|\bm{x}_{\sss\rm N}|}\!\ln\!{\frac{|\bm{x}_{\sss\rm N}|\!+\!\bm{n}\!\cdot\!\bm{x}_{\sss\rm N}}{|\bm{x}_{\rm e}|\!+\!\bm{n}\!\cdot\!\bm{x}_{\rm e}}} \!-\! \frac{1}{b^2}\!\ln\!{\frac{|\bm{x}_{\sss\rm N}|}{|\bm{x}_{\rm e}|}},\\
&&\int_{t_{\rm e}}^t \!\frac{\bm{n}\!\cdot\!\bm{x}_{\sss\rm N}}{|\bm{x}_{\sss\rm N}|^3} \!\ln\!{\frac{|\bm{x}_{\sss\rm N}|\!+\!\bm{n}\!\cdot\!\bm{x}_{\sss\rm N}}{|\bm{x}_{\rm e}|\!+\!\bm{n}\!\cdot\!\bm{x}_{\rm e}}} dt \!=\!-\frac{1}{|\bm{x}_{\sss\rm N}|} \!\ln\!{\frac{|\bm{x}_{\sss\rm N|}\!+\!\bm{n}\!\cdot\!\bm{x}_{\sss\rm N}}{|\bm{x}_{\rm e}|\!+\!\bm{n}\!\cdot\!\bm{x}_{\rm e}}} \!-\!\frac{1}{b}\!\bigg(\!\arccos\!{\frac{\bm{n}\!\cdot\!\bm{x}_{\sss\rm N}}{|\bm{x}_{\sss\rm N}|}} \!-\! \arccos\!{\frac{\bm{n}\!\cdot\!\bm{x}_{\rm e}}{|\bm{x}_{\rm e}|}}\!\bigg), \\
&&\int_{t_{\rm e}}^t \!\frac{\bm{n}\!\cdot\!\bm{x}_{\sss\rm N}}{|\bm{x}_{\sss\rm N}|^5} \!\ln\!{\frac{\!|\bm{x}_{\sss\rm N}|\!+\!\bm{n}\!\cdot\!\bm{x}_{\sss\rm N}\!}{|\bm{x}_{\rm e}|\!+\!\bm{n}\!\cdot\!\bm{x}_{\rm e}}}dt \!=\! -\frac{1}{3|\bm{x}_{\sss\rm N}|^3} \!\ln\!{\frac{\!|\bm{x}_{\sss\rm N}|\!+\!\bm{n}\!\cdot\!\bm{x}_{\sss\rm N}\!}{|\bm{x}_{\rm e}|\!+\!\bm{n}\!\cdot\!\bm{x}_{\rm e}}} \!-\!\frac{1}{6b^3}\!\bigg(\!\arccos\!{\frac{\!\bm{n}\!\cdot\!\bm{x}_{\sss\rm N}\!}{|\bm{x}_{\sss\rm N}|}} \!-\! \arccos\!{\frac{\!\bm{n}\!\cdot\!\bm{x}_{\rm e}\!}{|\bm{x}_{\rm e}|}}\!\bigg) \\
&& \hskip 4.75cm + \frac{1}{6b^2}\bigg(\!\frac{\!\bm{n}\!\cdot\!\bm{x}_{\sss\rm N}\!}{|\bm{x}_{\sss\rm N}|^2}\!-\!\frac{\!\bm{n}\!\cdot\!\bm{x}_{\rm e}\!}{|\bm{x}_{\rm e}|^2}\!\bigg)~.
\end{eqnarray*}

%
\end{document}